# Towards Power Efficient MAC Protocol for In-Body and On-Body Sensor Networks

Sana Ullah, Xizhi An, and Kyung Sup Kwak

Graduate School of Telecommunication Engineering
253 Yonghyun-Dong, Nam-Gu, 402-751, Inha University Incheon South Korea
sanajcs@hotmail.com, anxizhi@hotmail.com, kskwak@inha.ac.kr

**Abstract.** This paper presents an empirical discussion on the design and implementation of a power-efficient Medium Access Control (MAC) protocol for in-body and on-body sensor networks. We analyze the performance of a beacon-enabled IEEE 802.15.4, PB-TDMA, and S-MAC protocols for on-body sensor networks. We further present a Traffic Based Wakeup Mechanism that utilizes the traffic patterns of the BAN Nodes (BNs) to accommodate the entire BSN traffic. To enable a logical connection between different BNs working on different frequency bands, a method called Bridging function is proposed. The Bridging function integrates all BNs working on different bands into a complete BSN.

## 1 Introduction

The remote monitoring of body status and the surrounding environment are becoming more important for sporting activities, the safety of members of the emergency services, members of the military and health care. The levels of fitness required for the very competitive international sporting events require athletes to be at the very pinnacle of fitness with every muscle used to its utmost. Furthermore, many body functions are traditionally monitored only rarely and separated by a considerable period of time. This can give a very incomplete picture of what is really happening. Consider a patient visiting a doctor for a blood pressure check; he/she may be anxious and thus have elevated pressure resulting in an inaccurate diagnosis. If, however, the patient can be fitted with a simple monitoring system that requires no intervention, then a picture can be built up of how the pressure changes through the day when he/she goes about their normal business. This will give a better picture of what is happening and remove inaccurate results caused by going to visit the doctor. To achieve these requirements, monitoring of movement and body function are essential. This monitoring requires the sensors and wireless system to be very lightweight and to be integrated unobtrusively into the clothing.

   A Body Sensor Network (BSN) allows the integration of intelligent, miniaturized, low power, invasive and non-invasive sensor nodes to monitor body function and the surrounding environment. Each intelligent node has enough capability to process and forward information to a base station for diagnosis and prescription. A BSN provides long term health monitoring of patients under natural physiological states without constraining their normal activities. It can be used to develop a smart and affordable health care system and can be a part of diagnostic procedure, maintenance of chronic condition, supervised recovery from a surgical procedure and to handle emergency events [1].

A number of ongoing projects such as CodeBlue, MobiHealth, and iSIM have contributed to establish a proactive and unobtrusive BSN system [2]-[4]. A system architecture presented in [5] performs real time analysis of sensor's data, provides real time feedback to the user, and forwards the user's information to a telemedicine server. UbiMon aims to develop a smart and an affordable health care system [6]. MIT Media Lab is developing MIThril that gives a complete insight of human-machine interface [7]. HIT lab focuses on quality interfaces and innovative wearable computers [8]. IEEE 802.15.6 aims to provide power-efficient in-body and on-body wireless communication standards for medical and non-medical applications [9]. NASA is developing a wearable physiological monitoring system for astronauts called LifeGuard system [10]. ETRI focuses on the development of a low power MAC protocol for a BSN [11].

In this paper, we use the terms BAN Node (BN) and BAN Network Coordinator (BNC) for the sensor node and the network coordinator in a BSN. The rest of the paper is organized into six sections. Section 2 presents discussion on BSN traffic classification. Section 3 and 4 present a brief analysis on in-body and on-body MAC protocols. Section 5 and 6 discuss the Traffic Based Wakeup Mechanism and the Bridging function for a BSN. The final section concludes our work.

## 2  BSN Traffic Classification

The assorted BSN traffic requires sophisticated and power-efficient techniques to ensure safe and reliable operation. Existing MAC protocols such as SMAC [12], TMAC [13], IEEE 802.15.4 [14], and WiseMAC [15] give limited answers to the heterogeneous traffic. The in-body BNs do not appreciate synchronized wakeup periods because they confine the accommodation of sporadic emergency events. Medical data usually needs high priority and reliability than non-medical data. In case of emergency events, the BNs should be able to access the channel in less than one second [16]. IEEE 802.15.4 Guaranteed Time Slots (GTS) can be utilized to handle time critical events but they expire in case of a low traffic. Furthermore, some in-body BNs have high data transmission frequency than others. We classify the entire BSN traffic into Normal, On-demand, and Emergency traffic as given in Figure 1. The normal traffic is further classified into High, Medium, and Low traffic.

**(a)- Normal Traffic:** Normal traffic is the data traffic in a normal condition with no time critical and on-demand events. This includes unobtrusive and routine health monitoring of a patient for diagnosis and treatment of many diseases such as gastrointestinal tract, neurological disorders, cancer detection, handicap rehabilitation, and the most threatening heart disease. Some BNs have frequent wakeup periods and thus are designated as high traffic BNs. For example, an ECG node may send data 4 times per hour, while other BNs may send 4 times a day. The ECG node is thus designated as a high traffic BN. However, the normal traffic classification from high to low and vice-versa depends on the application requirements. The normal data is collected and processed by the BNC.

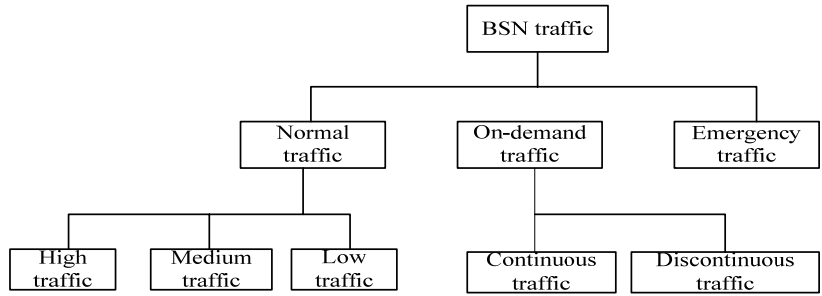

**Fig. 1.** BSN Traffic Classification

**(b)- On-demand Traffic:** On-demand traffic is initiated by the BNC or doctor to know certain information, mostly for the purpose of diagnosis and prescription. This is further divided into continuous (in case of surgical events) and non-continuous (when limited information is required).

**(c)- Emergency Traffic:** This is initiated by BNs when they exceed a predefined threshold and should be accommodated in less than one second. This kind of traffic is not generated on regular intervals and is totally unpredictable.

## 3   MAC for On-Body Sensor Networks

On-body sensor networks comprise of miniaturized and non-invasive sensor nodes that are used for various applications, ranging from medical to interactive gaming and entertainment applications. They use Wireless Medical Telemetry Services (WMTS), unlicensed ISM, and UWB bands for data transmission. WMTS is a licensed band designated for medical telemetry system. Federal Communication Commission (FCC) urges the use of WMTS for medical applications due to fewer interfering sources. However, only authorized users such as physicians and trained technicians are eligible to use this band. Furthermore, the restricted WMTS (14 MHz) bandwidth cannot support video and voice transmission. The alternative spectrum for medical applications is to use 2.4 GHz ISM band that includes guard bands to protect adjacent channel interference.

The design and implementation of a power-efficient MAC protocol for on-body sensor networks have been a hot research topic for the last few years. A novel TDMA protocol for on-body sensor network called H-MAC exploits the biosignal features to perform TDMA synchronization and improves the energy efficiency [17]. Other protocols like WASP, CICADA, and BSN-MAC are investigated in [18] – [20]. The performance analysis of a non-beacon IEEE 802.15.4 is adapted to extend lifetime of a node from 10 to 15 years [21]. This work considers low upload/download rates, mostly per hour. Furthermore, the data transmission is based on periodic intervals, which limits the performance to certain applications. There is no reliable support for on-demand and emergency traffic.

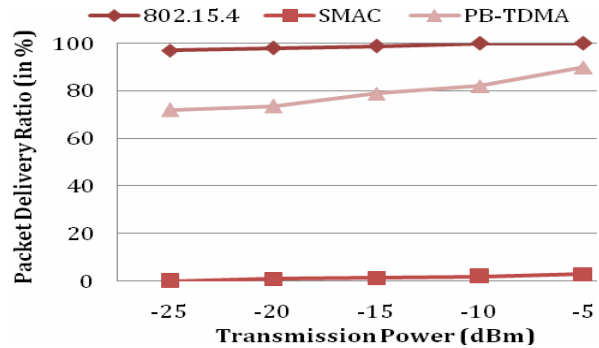

**Fig. 2.** Packet Delivery Ratio of IEEE 802.15.4, PB-TDMA, and SMAC for On-Body Sensor Networks

We analyze the performance of a beacon-enabled IEEE 802.15.4, Preamble-Based TDMA [22], and SMAC protocols for on-body sensor networks. Our analysis is verified by extensive simulation using NS-2[23]. In case of S-MAC, PB-TDMA, and IEEE 802.11 (DCF) protocols, the wireless physical parameters are considered according to low power Nordic nRF2401 transceiver. This radio transceiver operates in the 2.4-2.5 GHz band with an optimum transmission power of -5dBm. However, in case of IEEE 802.15.4, Chipcon CC2420 radio interface is considered. We use shadowing propagation model throughout the simulations. We consider 9 BNs firmly placed on a human body. The BNs are connected to the BNC in a star topology. The initial BN energy is 5 Joules. The data rate of the BNs is heterogeneous. The simulation area is 1x1 meter and each BN generates Constant Bit Rate (CBR) traffic. The packet size is 128 bytes. The transport agent is User Datagram Protocol (UDP). Simulation results show that IEEE 802.15.4, when configured in a beacon-enabled mode, outperforms SMAC and PB-TDMA as shown in Figure 2. However, the precise location of BNs and the body position influence the packet delivery ratio.

Intel Corporation conducted a series of experiments to analyze the performance of IEEE 802.15.4 for an on-body sensor network [24]. They deployed a number of Intel Mote 2 [25] nodes on chest, waist, and the right ankle. Tabel 1 shows the packet success rate at 0dBm transmit power when a person is standing and sitting on a chair. The connection between ankle and waist cannot be established, even for a short distance of 1.5m. All other connections show favourable performance.

As IEEE 802.15.4 operates in the 2.4 GHz unlicensed band, the possibilities of interference from other devices such as IEEE 802.11 and microwave are inevitable. A series of experiments to evaluate the impact of IEEE 802.11 and microwave ovens on the IEEE 802.15.4 transmission are carried out in [26]. They considered XBee 802.15.4 development kit that has two XBee modules. Table 2 shows the effects of microwave oven on the XBee remote module. When the microwave oven is ON, the packet success rate and the standard deviation is degraded to 96.85% and 3.22% respectively. However, there is no loss when the XBee modules are taken 2 meters away from the microwave oven.

**Table 1.** Packet Success Rate at 0dBm Transmit Power

| | Packet Success Rate when a Person is Standing | | | Packet Success Rate when a Person is Sitting on an Office Chair | | |
|---|---|---|---|---|---|---|
| Source \ Destination | Chest | Waist | Ankle | Chest | Waist | Ankle |
| Chest | | 99% | 84% | | 99% | 81% |
| Waist | 100% | | 50% | 99% | | 47% |
| Ankle | 72% | 76% | | 77% | 27% | |

**Table 2.** Co-existence Test Results between IEEE 802.15.4 and Microwave Oven

| Microwave Status | Packet Success Rate | |
|---|---|---|
| | Mean | Std. |
| ON | 96.85% | 3.22% |
| OFF | 100% | 0% |

## 4   MAC for In-Body Sensor Networks

The most challenging task in developing a power-efficient MAC for in-body sensor networks is to accommodate the in-body BNs in a power-efficient manner. Unlike on-body BNs, the in-body BNs are implanted under human skin where the electrical properties of the body affect the signal propagations. The human body is a medium that poses many wireless transmission challenges. The body is composed of several components that are not predictable and will change. Monitoring in-body functions and the ability to communicate with implanted therapeutic devices, such as pacemakers, are essential for their best use.

   Zarlink semiconductor has introduced a wireless chip that supports a very high data rate RF link for communication with an implantable BN [27]. The ZL70101 ultra-low power transceiver chip satisfies the power and size requirements for implanted communication systems and operates in 402-405 MHz MICS band [28]-[29]. Other frequency bands such as 916MHz, 1.5GHz, and UWB are also considered for in-body sensor networks [30]-[32]. The use of open air models for implant communication is discouraged in [33]. A Finite-Difference Time-Domain (FDTD) is used to calculate the power deposition in a human head and is measured by the SAR in W/Kg [34]. However, the distance from the implant source has not been discussed. For a duty cycle and transmission rate of 0.05% and 400 kbps, the SubQore radio architecture from Cambridge university consumes a peak current less than 1.7mA [35]. An implantable medical microsystems for interfaces to the central nervous system is presented in [36].

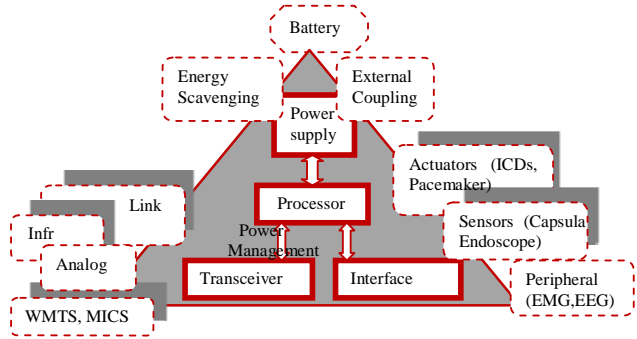

**Fig. 3.** Implementation issues in the in-body sensor network

The diverse nature of in-body BNs together with the electrical properties of the human body influences the development process of a power-efficient MAC for in-body sensor networks. The data rate of implanted BNs varies, ranging from few kbps in pacemaker to several Mbps in capsular endoscope. Figure 3 explains the in-body sensor networks and implementation issues. In the in-body sensor network, critical traffic requires low latency and high reliability than non-critical traffic. One of the solutions is to adjust initial back-off windows in a traditional CSMA/CA for critical and non-critical traffic. Due to high path loss inside the human body, the use of CSMA/CA does not provide reliable solution in multi-piconets [37]. For a threshold of -85dBm and -95dBm, the on-body BNs cannot see the activity of in-body BNs when they are away at 3 meters distance from the body surface. However, within 3 meters or less distance, the CCA works correctly in the same piconet.

The in-body MAC should also consider the thermal influence caused by the electromagnetic wave exposure and circuit heat. Nagamine *et al.* discussed the thermal influence of the BNs using different MAC protocols [38]. Figure 4 shows the temperature of a BN when ALOHA and CSMA/CA are used.

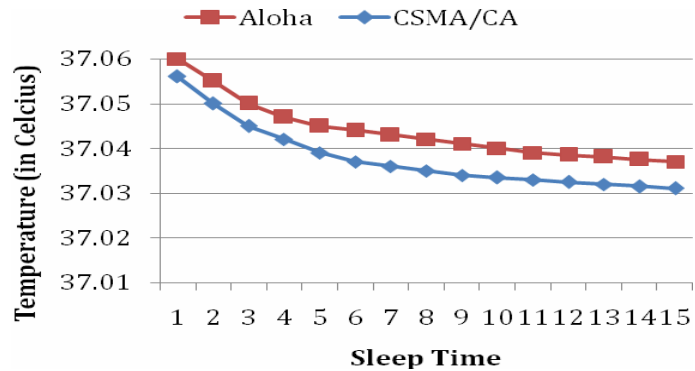

**Fig. 4.** Saturated Temperature

## 5   Traffic Based Wakeup Mechanism for a BSN

The heterogeneous BNs require power-efficient and dynamic wakeup techniques for reliable operation. We propose a power-efficient technique called Traffic Based Wakeup Mechanism for a BSN that exploits the traffic patterns of the BNs to accommodate the assorted BSN traffic. The initial wakeup patterns are either predefined (by the company) or created and modified (by the BNC). Table 3 shows the traffic classification and the corresponding solutions.

**Table 3.** Solutions to the Classified BSN traffic

| Traffic / Devices | Normal Traffic | | | On-demand Traffic | Emergency Traffic |
|---|---|---|---|---|---|
| | High | Medium | Low | | |
| BAN Nodes (BNs) | Send data based on the **Traffic-based Wake-up Table** | | | Receives a **Wakeup Radio Signal** from the BNC and respond | Send a **Wake-up Radio Signal** to the BNC in case of emergency |
| BAN Coordinator (BNC) | Send beacons based on the **Traffic- based Wake-up Table** | | | Send a **Wake-up Radio Signal** to BNs | Receives a **Wake-up Radio Signal** and respond |

The wakeup patterns of all BNs are organized into a table called Traffic Based Wakeup Table. The table is maintained and modified by the BNC according to the application requirements. Based on the BNs wakeup patterns, the BNC can also calculate its own wakeup pattern. This could save significant energy at the BNC. The BNC does not need to stay active when there is no traffic from the BN. The designation of normal traffic levels, i.e, high, medium, and low traffic nodes depend on the application. For emergency and on-demand traffic, the BNs and the BNC send a wake-up signal for a very short duration to each other. However, traditional wake-up radio concepts have several limitations when considered in the in-body sensor networks. They are not able to wake-up a particular BN. All BNs wake-up in response to a single wake-up signal, which is not the required environment. The use of different radio frequencies to wake-up a particular BN may provide an optimal solution. However, the use of wakeup radio requires efficient security management schemes.

## 6   Bridging Function for a BSN

In a BSN, there can be various BNs working on different frequency bands and have correspondingly different Physical Layers (PHYs). The main problem is how to connect different BNs working on different bands in a BSN. In order to accommodate multiple PHYs (radio interfaces) and multiple channels, we introduce a function called Bridging that virtually connects different BNs working on different PHYs.

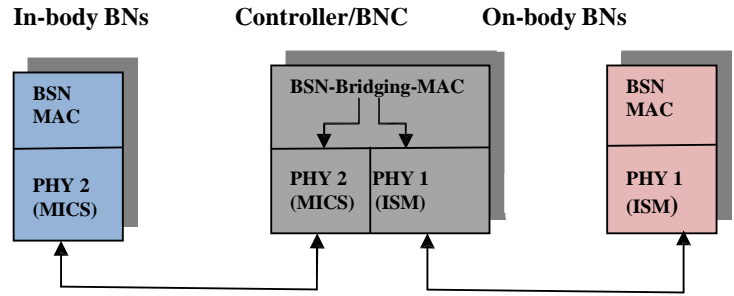

**Fig. 5.** Protocol Stack of the Bridging Function

Bridging function enables one MAC to support multiple bands and multiple PHYs. One MAC means a common hybrid MAC framework. Furthermore, we use the term Channel from MAC point of view where one channel is a combination of a frequency band and the corresponding PHY technique.

The Bridging function is an enhanced MAC function that establishes logical relationships between different PHY BNs. However, the BNs running Bridging function must have two or more different PHY interfaces in order to work on two or more different Channels at the same time. All necessary information, i.e, Network Info, Channel ID (band/PHY), BNs ID, Connection ID and Connection Type (MAC layer), Source and Destination BNs ID is recorded in a specific table. According to the records in the table, a Channel Mapping is implemented in the intermediate BNs that support at least two bands/PHYs.

The Bridging function is also regarded as link-layer MAC Protocol Data Unit (MPDU) relay. The intermediate BNs receive and store data from one channel, and then forward it on another channel towards the destination. The process is transparent to upper layers and accommodates different PHY techniques. Figure 5 shows the protocol stack of the Bridging function.

The BNs implementing the Bridging function (also called a Bridge) can collect or dissipate data from or to the in-body BNs and communicate with on-body and out-body BNs. The Bridge has two PHY layers, typically unlicensed ISM/UWB Band and a licensed MICS Band. Generally, on-body/out-body BNs are selected for the Bridging function due to their relative larger capabilities and less stringent channel conditions. The Bridging enables the integration of all BNs into a BSN and realizes the interconnectivity among different PHY BNs.

The in-body BNs cannot establish a direct communication with on-body BNs due to power limitations, high path loss, and different PHYs. Therefore, they exploit the Bridging function and forward all the data frames to the on-body BNs via Bridge. Moreover, the in-body BNs do not support peer-to-peer transmission in same network, and thus data transmission between in-body BNs are relayed by the Bridge. Though on-body/out-body BNs can be designated as a Bridge node, but we urge the use of BNC to perform the Bridging function.

## 7 Conclusions

In this paper, we studied the behavior of several power-efficient MAC protocols including a beacon-enabled IEEE 802.15.4 protocol for on-body sensor networks. We concluded that none of the existing MAC protocols accommodate the assorted BSN traffic requirements in a power-efficient manner. The entire BSN traffic was classified into normal, on-demand, and emergency traffic. A Traffic Based Wakeup Mechanism was proposed for a BSN, which exploited the traffic patterns of the BNs to accommodate the entire traffic classification. We further introduced a Bridging function that integrated all the BNs working on different PHYs into a complete BSN. The proposed wakeup mechanism backed by the Bridging function provided a complete solution towards power-efficient and reliable communication in a BSN.

## Acknowledgement

This work was supported by the IT R&D program of MKE/IITA, [2008-F-050-01, Development of WBAN system for In-body and On-body].